\documentclass{elsart}
\usepackage{amssymb}
\usepackage[german,english]{babel}
\usepackage[intlimits,sumlimits,namelimits]{amsmath}
\usepackage[dvips]{epsfig}
\usepackage[dvips]{color}
\usepackage{ifthen,shadow}
\usepackage{fancyhdr}

\newcommand{\beqa}{\begin{eqnarray}}
\newcommand{\eeqa}{\end{eqnarray}}
\begin{document}

\begin{frontmatter}

\title{\bf Extracting a value of the slope of the neutron form factor 
$G_{En}(q^2)$ at $q^2=0$ by reanalysis of the experimental data}

\author{M.M.\ Mustafa, E.M.\ Darwish, E.A.M.\ Sultan}

\address{Physics Department, Faculty of Science, South Valley University, 
Sohag, Egypt}

\author{E.M.\ Hassan}

\address{Physics Department, Faculty of Science, South Valley University, 
Qena, Egypt}

\date{\today}

\begin{abstract}
A new value $b=0.0205\pm0.0017$ fm$^2$ of the slope of the neutron 
form factor $G_{En}(q^2)$ at $q^2=0$ compatible with deuteron properties has 
been extracted by using a linear relation between $b$ and $A^2_S(1+\eta^2)$ we 
found for a class of nonlocal potential models having the experimental values 
of both $r_D$ and $Q$. Another {\it model dependent} value 
$b^{\rm MHKZ}=0.0206\pm0.0014$ fm$^2$, which is also compatible with deuteron 
properties, has been determined by applying "constrained" unitary 
transformations to the local MHKZ potential model. The sensitivity of a small 
changes in the experimental values used for $r_D$ and $Q$ on the value 
obtained for $b$ is also investigated.

\vspace{0.5cm}

\end{abstract}
\end{frontmatter}

\section{Introduction} 
\label{sec1}
Measurements of the elastic scattering of electrons by deuterons lead in a
direct way to the deuteron form factors as functions of the momentum
transfer $q^2$. These form factors are of interest in 
connection with the study of the structure of the deuteron and the 
determination of the electromagnetic form factors of the neutron \cite{1}.

Correlations between the slope $b$ of the neutron electric form factor 
$G_{En}(q^2)$ at ${q^2=0}$ and the deuteron root mean square (rms) matter 
radius $r_D$ for different NN-potential models have been used to determine 
$r_D$ \cite{2,3,4}. Berard {\it{et al.}} \cite{2} found a linear relation 
with a positive 
slope between $b$ and $r_D^2$ for a class of potential models. 
They used their experimental data for the ratio $R(q^2)$ of the deuteron to 
the proton electric form factors in the range of momentum transfers 
$0.05\le q^2\le 0.50$ fm$^{-2}$ and the experimental value 
$b=0.0189\pm0.0004$ fm$^2$ given by Krohn and Ringo \cite{5} to extract the value 
$r_D=1.9635\pm0.0045$ fm. Allen {\it{et al.}} \cite{3} found a linear relation with 
a positive slope between $b$ and $r_D^2$ for the radial wave functions for a 
single local potential model and its family of the wave functions produced by 
unitary transformations. The experimental data of $R(q^2)$ of Berard 
{\it{et al.}} \cite{2} and the experimental value $b=0.0199\pm0.0003$ fm$^2$ given 
by Koester {\it{et al.}} \cite{6} have been used to determine the model dependent 
value $r_D=1.952\pm0.004$ fm and the model independent one $r_D=1.948\pm0.023$ 
fm indirectly from the plotted straight line. They used also the experimental 
data of $R(q^2)$ of Akimov {\it{et al.}} \cite{1} in the range 
$0.05\le q^2\le 0.50$ fm$^{-2}$ to determine $r_D=2.005\pm0.118$ fm. In an 
analogous procedure Mustafa {\it{et al.}} \cite{4} found a linear relation between 
$b$ and $r_D^2$ for the radial wave functions for the MHKZ local potential 
model \cite{4} and its family of the wave functions produced by unitary 
transformations. They used also the experimental data of $R(q^2)$ of Berard 
{\it{et al.}} \cite{2} and the experimental value $b=0.0199\pm0.0003$ fm$^2$ given 
by Koester {\it{et al.}} \cite{6} to determine $r_D=1.9546\pm0.0021$ fm.

In this work, correlations between $b$ and deuteron properties, i.e. 
the asymptotic S-state amplitude $A_S$, the asymptotic D/S ratio 
$\eta$, the rms radius $r_D$, the quadrupole moment $Q$ and the 
binding energy $E_b$, have been used to determine $b$. Two linear relations 
between $b$ and $A_S^2(1+\eta^2)$ have been found for deuteron potential 
models, one with a positive slope and the other with a negative one. 

At large distances, outside the nuclear potential range, the $^3S_1$- and 
the $^3D_1$-state radial wave functions ($u$ and $w$, respectively) behave 
asymptotically as \cite{7}
\beqa
u(r) & = & A_S~e^{-\gamma r}\,,\nonumber \\
\nonumber \\
w(r) & = & A_D\biggl(1+\frac{3}{\gamma r}+\frac{3}{(\gamma r)^2}\biggr)
e^{-\gamma r}\,,
\label{eq1}
\eeqa
where $\gamma^2=-\frac{2m}{\hbar^2}E_b$ \cite{8} and $m$ is the reduced 
neutron-proton mass. If the proton and the neutron are at a distance $r$ 
apart, the deuteron rms radius $r_D$ may be written 
as \cite{7}
\beqa
r_D^2 & = & \frac{1}{4}\int_{0}^{\infty}r^2~(u^2+w^2)dr\,.
\label{eq2}
\eeqa
The deuteron quadrupole moment $Q$ may be also given by \cite{7}
\beqa
Q & = & \frac{1}{\sqrt{50}}\int_{0}^{\infty}r^2uwdr-\frac{1}{20}\int_{0}^{\infty}
r^2w^2dr\,.
\label{eq3}
\eeqa
The correlation with $A^2_S(1+\eta^2)$ has been used 
because the term $(u^2+w^2)$ which has been involved in the definition of 
$r_D^2$ proportional to $A^2_S(1+\eta^2)$ if we use the asymptotic forms of 
the radial  wave functions $u$ and $w$ of Eq.~(\ref{eq1}) and ignore the centrifugal 
term $[1+3/(\gamma r)+3/(\gamma^2 r^2)]$.

The aim of this work is to extract a new value of $b$ which is compatible 
with deuteron properties. This value is extracted indirectly 
by using the linear correlation between $b$ and $A_S^2(1+\eta^2)$ of nonlocal 
potential models (having the experimental values of both $r_D$ and $Q$) 
produced by applying unitary transformations to standard nonrelativistic 
potential models.

In Sect.~\ref{sec2} we will outline the method which has been used to calculate a 
model value of the slope $b$ of the neutron form factor $G_{En}(q^2)$ at 
$q^2=0$. The new determinations of this parameter using the local and nonlocal 
potential models will be distinguished in Sect.~\ref{sec3}. In Sect.~\ref{sec4} a simple 
method to calculate the parameter $b$ using one single potential model will 
be obtained. Finally, the sensitivity of a small changes in the experimental 
values used for $r_D$ and $Q$ on the extracted value of $b$, i.e. 
$\frac{\delta b}{\delta r_D}$ and $\frac{\delta b}{\delta Q}$, will be given 
in Sect.~\ref{sec5}.

\section{Calculating model values of $b$}
\label{sec2}
In order to extract the slope $b$ of the neutron form factor $G_{En}(q^2)$ 
at $q^2=0$ from elastic electron-deuteron scattering, we start from the 
expression of the deuteron electric form factor 
in the nonrelativistic impulse approximation, neglecting two-body 
contributions \cite{7}
\beqa
G_{ED}(q^2) & = & \biggl[G_{Ep}(q^2)+G_{En}(q^2)\biggr]\biggl[
C_E^2(q^2)+C_Q^2(q^2)\biggr] ^{1/2}\biggl(1+\tau\biggr)^{-1/2}\nonumber\\
\nonumber \\ 
 & \simeq & \biggl[G_{Ep}(q^2)+G_{En}(q^2)\biggr]C_E(q^2)\biggl(1+\tau\biggr) 
^{-1/2}\,,
\label{eq4}
\eeqa
where $G_{Ep}(q^2)$ is the proton electric form factor, $C_E(q^2)$ and 
$C_Q(q^2)$ are electric charge and quadrupole form factors, 
respectively, which are given in terms of the nonrelativistic S- and D-wave 
functions by \cite{7}
\beqa
C_E(q^2) & = & \int_{0}^{\infty}\left(
u^2+w^2\right)~j_0\left(\frac{qr}{2}\right)dr
\label{eq5}
\eeqa
and
\beqa
C_Q(q^2) & = & 2\int_{0}^{\infty}\left(uw-\frac {w^2}{\sqrt
8}\right)~j_2\left(\frac{qr}{2}\right)dr\,,
\label{eq6}
\eeqa
where $j_0$ and $j_2$ are the spherical Bessel functions of order zero and
two, respectively. The factor $\tau=(q^2/4m_p^2)$ is called the Darwin-Foldy
correction to the nucleon form factors \cite{9}, where $m_p=4.5749098$ fm$^{-1}$ is 
the proton mass. Note that, Eq.~(\ref{eq4}) is usually used because $C_Q\ll C_E$ for 
small $q^2$.

The value of $b$ for a particular potential model is insensitive to the choice 
of the proton electric form factor $G_{Ep}(q^2)$. In this work we take the 
parameterization of Simon {\it{et al.}} \cite{10} for $q^2$ in fm$^{-2}$
\beqa
G_{Ep}(q^2) & = & \frac{0.312}{\left( 1+\frac{q^2}{6}\right)
}+\frac{1.312}{\left( 1+\frac{q^2}{15.02}\right)
}-\frac{0.709}{\left( 1+\frac{q^2}{44.08}\right)
}+\frac{0.085}{\left( 1+\frac{q^2}{154.2}\right) }\,.
\label{eq7}
\eeqa

For the evaluation of the charge form factor $C_E(q^2)$ given in Eq.~(\ref{eq5}) we 
split the integration according to 
\beqa
C_E(q^2) & = & \int_{0}^{R}\left( u^2+w^2\right) j_0\left(\frac{qr}{2}\right)dr+\Delta C_E(q^2)\,,
\label{eq8}
\eeqa
where $\Delta C_E(q^2)$ is the asymptotic analytic correction to $C_E(q^2)$
given by
\beqa
\Delta C_E(q^2) & = & \int_{R}^{\infty}\left( u^2+w^2\right)
j_0\left(\frac{qr}{2}\right)dr\,.
\label{eq9}
\eeqa

The first integration of Eq.~(\ref{eq8}) is calculated numerically from $r=r_c$ to 
$r=R$, where $r_c$ is the hard-core radius - if any - and $R$ is chosen to 
be 16+$r_c$ fm which is beyond the range of the nuclear potential. The 
integral in Eq.~(\ref{eq9}) is carried out analytically from $r=R$ to $r=\infty$ by 
using the asymptotic forms of the radial wave functions $u$ and $w$ of Eq.~(\ref{eq1}). 
The analytic formula of the $\Delta C_E(q^2)$ is given in Ref.\cite{4}, for which 
the values of $C_E(q^2)$ are correct in the momentum transferred range 
$0\le q^2 \le 25$ fm$^{-2}$. In an analogous procedure the deuteron rms 
radius $r_D$ given in Eq.~(\ref{eq2}) and the deuteron quadrupole moment $Q$ given in 
Eq.~(\ref{eq3}) are evaluated. 

Given $C_E(q^2)$ for a particular potential model, then the experimental 
values of the ratio $R(q^2)$ of the deuteron form factor $G_{ED}(q^2)$ to 
the proton form factor $G_{Ep}(q^2)$ can be used to determine the neutron 
form factor \cite{2} from
\beqa
G_{En}(q^2) & = & \left[\frac {R(q^2)\sqrt {1+\tau}}{C_E(q^2)}-1\right]G_{Ep}(q^2)
\label{eq10}
\eeqa
and hence, the slope $b$ of $G_{En}(q^2)$ at $q^2=0$, 
\beqa
b & = & \left.\frac {dG_{En}(q^2)}{dq^2}\right\vert _{q^2=0}\,.
\label{eq11}
\eeqa
The experimental values of $R(q^2)$ of Simon {\it{et al.}} \cite{11} are used in 
this work. This set of data covers the range of values 
$0.044\le q^2\le 4$ fm$^{-2}$.

Various parameterizations have been used for $G_{En}(q^2)$ \cite{2,12,13,14,15,16,17,18}. In 
order to calculate $b$ in this work we parameterize $G_{En}(q^2)$ by a 
polynomial of order three in $q^2$ as
\beqa
G_{En}(q^2) & = & bq^2+cq^4+dq^6
\label{eq12}
\eeqa
which is agrees well for $0.044\le q^2\le 4$ fm$^{-2}$. The model dependency 
of $b$, here, is a result of the model dependence of the 
charge form factor $C_E(q^2)$ via the radial wave functions.

\section{The new determinations of $b$}
\label{sec3}
In order to extract a new value of the slope $b$ of the neutron form factor 
$G_{En}(q^2)$ at $q^2=0$, a linear relation between $b$ and $A_S^2(1+\eta^2)$ 
with a positive slope has been found for standard nonrelativistic potential 
models as shown in Fig.~\ref{fig1}. These models and the "names" given to 
them are the potentials of Glendenning and Kramer "GK1, ....., GK9" \cite{19}, 
Lacombe {\it{et al.}} "PARIS" \cite{20}, Mustafa {\it{et al.}} "MHKZ" \cite{4}, 
Reid "RHC, RSC, RSCA" \cite{21}, Machledit {\it{et al.}} "MACH-A, -B, -C" \cite{22}, 
Machledit {\it{et al.}} "Bonn-F, -Q" \cite{23}, de Tourreil and Sprung "TS-A, -B, 
-C" \cite{24}, de Tourreil {\it{et al.}} "TRS" \cite{25}, Hamada and Johnston "HJ" \cite{26}, 
Mustafa and Zahran "MZ" \cite{27}, Mustafa "A, B" \cite{28}, Mustafa {\it{et al.}} 
"r1, r3, ...., r7" \cite{29}, Mustafa "L1, L2, 1, 2, ....., 6" \cite{30} and Mustafa 
"a, b, ......, i" \cite{31}. The value
\beqa
b & = & 0.0296\pm0.0118~ {\rm fm}^2
\label{eq13}
\eeqa
which is compatible with the experimental values of $A_S$, $\eta$ and $E_b$ 
but not with the experimental values of $r_D$ and $Q$ is extracted from this 
correlation. It is the value of $b$ corresponding to the experimental value 
$A_S^2(1+\eta^2)=0.7817$ fm$^{-1}$. This value of $b$ is greater than the 
previously extracted value $b=0.0199\pm0.0003$ fm$^2$ by Koester 
{\it{et al.}} \cite{6} and its error $\Delta b$ is relatively large. 
Therefore, the result obtained for $b$ using the local potential models is 
not very reliable.

Moreover, it is desirable to obtain a value of $b$ which is not only 
compatible with the experimental values of $A_S$, $\eta$ and 
$E_b$, but also with the experimental values of $Q$ and $r_D$, which have 
dependencies on the interior parts of the radial wave functions. 
The experimental values of these quantities are listed in Table \ref{tab1}. 
Contributions to the value of $r_D$ and $Q$ from meson exchange currents 
and other relativistic corrections $\Delta r_D=0.0034$ fm and 
$\Delta Q=0.0063$ fm$^2$ of Kohno \cite{35} are neglected in the present work. 
Kermode {\it{et al.}} \cite{36} and Mustafa and Hassan \cite{37} have proven that $Q$ 
and $r_D$ are not purely asymptotic quantities by showing that 
phase-equivalent potentials can have different quadrupole moments and radii. 
Therefore, the linear relation between $b$ and $A_S^2(1+\eta^2)$ obtained in Fig.~\ref{fig1}  has been re-drawn using only potential models which 
reproduce the experimental values of both $Q$ and $r_D$. These phase-equivalent potential models 
are produced by using "constrained" short-range unitary transformations like those used previously by Kermode {\it{et al.}} \cite{36} for 
the radial wave functions ($u_i,w_i$) of a local potential models. The transformed wave functions ($\overline u_i,\overline w_i$) of a nonlocal potential 
model are given by
\beqa
\overline u_i & = & u_i-2g(r)\int_0^\infty g(s) u_i(s) ds\,,\nonumber \\
\nonumber \\
\overline w_i & = & w_i-2g(r)\int_0^\infty g(s) w_i(s) ds\,.
\label{eq14}
\eeqa
Various parameterizations have been used for the function $g(s)$ \cite{37,38,39,40,41}. 
We have chosen the parameterization $g(s)=Cs(1-\beta s)e^{-\alpha s}$, where 
$s=r-r_c$ and $C=[4\alpha^5/(\alpha^2-3\beta\alpha+3\beta^2)]^{1/2}$~ is a 
normalizing factor such that $\int_{0}^{\infty}g^2 (r)dr=1$. The parameters 
$\alpha$ and $\beta$ are adjustable parameters. The parameter $\alpha$ can 
be regarded as the "range" and $\beta$ as the "strength" of the unitary 
transformation. The values of $\alpha < 1.5$ fm$^{-1}$ produce wave 
functions of different asymptotic behaviour \cite{37}. In this work, the parameter 
$\alpha$ is assumed to changed from $\alpha=1.5$ fm$^{-1}$ to 
$\alpha=7$ fm$^{-1}$ in steps of 0.0001 fm$^{-1}$. For each value of $\alpha$, 
the parameter $\beta$ is changed from $\beta=0$ fm$^{-1}$ to 
$\beta=4$ fm$^{-1}$ in steps of 0.0001 fm$^{-1}$. The function $g(r)$ is of 
short range, hence, the wave 
functions $u$ and $w$ of a local potential model and $\overline u$ and 
$\overline w$ of the corresponding nonlocal potential model are the same 
in the asymptotic region. This implies that the asymptotic quantities $A_S$ 
and $A_D$ and hence, $\eta=A_D/A_S$, are the same for both the local and 
nonlocal potential models, but $r_D$ and $Q$ could be different.

The values of the non-locality parameters $\alpha$ and $\beta$ are adjusted 
to produce transformed wave functions having the experimental values 
of $r_D$ and $Q$. This could not be achieved for all the local potentials 
considered. It is always possible to find for a given value of $\alpha$ a value of $\beta$ which 
corresponds either to the experimental value of $r_D$ or to the experimental 
value of $Q$. In fact, for a given $\alpha$ one often finds two solutions for 
$\beta$ as is shown by the two branches in Fig.~\ref{fig2}. But it was not always 
possible for some potential models to find pairs ($\alpha,\beta$) such that 
both $r_D$ and $Q$ could be reproduced. In this case we fix the 
non-locality parameters ($\alpha,\beta$) which give the experimental value of 
$r_D$ by searching for the closest value of $Q$ with respect to the 
experimental value or vice versa. If both values would be fitted, the two type 
of curves possess an intersection point as is demonstrated in Fig.~\ref{fig2} for 
selected potential models. The transformed wave functions having the 
experimental values of both $r_D$ and $Q$ are compared to the radial wave 
functions in Fig.~\ref{fig3} for selected potential models. Altogether, we have studied 
forty-nine potential models. Out of these, we only found seventeen transformed models giving the 
experimental values for both $r_D$ and $Q$. The pairs of the non-locality 
parameters ($\alpha,\beta$) having the experimental values of both $r_D$ and $Q$ 
are listed in Table \ref{tab2}.

In this case of applying ``constrained'' unitary transformations, a linear 
relation with a negative slope between $b$ and $A_S^2 (1+\eta^2)$ is found 
as shown in Fig.~\ref{fig4}. The points representing the transformed potential models 
lie on or closely scattered around the straight line. We would like to 
mention that the five points representing the nonlocal potentials which are 
phase equivalent to the family of the Glendenning and Kramer potential models 
(GK2, GK3, GK5, GK7 and GK8) \cite{19} lie on a separate line with a different 
negative slope (see also, Fig.1 in Klarsfeld {\it{et al.}} \cite{42} in which 
they used deuteron potential models in an empirical relation and found that 
the point representing the Glendenning and Kramer potentials also not lie 
on the empirical line, but lie on a separate line with a different slope). 
However, these old models predict a relatively larger deuteron 
binding energy $E_b$ in agreement with the then experimental value 
$E_b=-2.226\pm0.003$ MeV \cite{43}. Therefore, these five potential models are not 
considered in Fig.~\ref{fig4}. The new value 
\beqa
b & = & 0.0205\pm0.0017~{\rm fm}^2
\label{eq15}
\eeqa
is extracted from the straight line of Fig.~\ref{fig4} using only the remaining twelve 
nonlocal potential models. This value of $b$ is compatible with the 
experimental values of $A_S$, $\eta$, $r_D$, $Q$ and $E_b$. It is more 
consistent with the previously published values $b=0.0189\pm0.0004$ fm$^2$ of 
Krohn and Ringo \cite{5} and, $b=0.0199\pm0.0003$ fm$^2$ of Koester \etal \cite{6}. 
The standard error in Eq.~(\ref{eq15}) is much less than that of Eq.~(\ref{eq13}), but is 
greater than those of the previous measurements.

\section{Determination of $b$ using MHKZ potential model}
\label{sec4}
So far, we have used two methods to extract the slope parameter $b$. The first 
uses the linear relation between $b$ and $A_S^2(1+\eta^2)$ of the local 
potential models as shown in Fig.~\ref{fig1}. The second is based on Fig.~\ref{fig4} using the nonlocal potential 
models having the experimental values of both $r_D$ and $Q$. The two extracted 
values in Eqs.(13) and (15) differ substantially.

A third method to extract $b$ does not use the linear relation between $b$ and $A_S^2(1+\eta^2)$ and a large number of potential models. It uses only one {\it{single}} potential 
model which is the potential model of Mustafa {\it{et~al.}} \cite{4}. The 
importance of the MHKZ
potential model comes from the fact that it reproduces the experimental values of $A_S$ and $\eta$.
The unitary transformation applied to the MHKZ wave functions are 
constrained to produce a phase-equivalent transformed wave functions 
having the experimental values of both $r_D$ and $Q$. The {\it 
model} value,
\beqa
b^{\rm MHKZ} & = & 0.0206\pm0.0014~{\rm fm}^2
\label{eq16}
\eeqa
is extracted by using the phase-equivalent potential of the MHKZ potential 
model \cite{4}. It is also compatible with the 
experimental values of $A_S$, $\eta$, $E_b$, $r_D$ and $Q$. This value is in 
very good agreement with the value obtained in Eq.~(\ref{eq15}) of this work.

\section{The sensitivity of a small changes in the experimental values used 
for $r_D$ and $Q$ on the extracted value of $b$}
\label{sec5}
The dependence of the determined value of $b$ on the experimental values used 
for $r_D$ and $Q$ has been investigated. The whole procedure has been repeated 
twice, first by allowing for a small change $\delta r_D$ in the value assumed 
as an experimental value of $r_D$ and fixing the experimental values of $A_S$, 
$\eta$, $E_b$ and $Q$ to find the corresponding change $\delta b$. 
The published experimental values of $r_D$ are in the range between 
$r_D=1.947\pm0.029$ fm of Akimov \etal \cite{1} and $r_D=1.9635\pm0.0045$ fm of 
Berard \etal \cite{2}. The value of $r_D$ of Akimov \etal \cite{1}, which is 
the lower limit of the experimental determination of $r_D$, has been 
used in this work to study the correlation between the determined value 
of $b$, Eq.~(\ref{eq15}), and the value used for $r_D$ as an experimental value. 
The value
\beqa
b & = & 0.0176\pm0.0013~{\rm fm}^2
\label{eq17}
\eeqa
is extracted in this case by using the transformed wave functions 
of potential models having the values of both $r_D$ and $Q$. 
We show that the decrease in the value of $r_D$ leads to a corresponding 
decrease in the value of $b$. Therefore, we find
\beqa
\frac{\delta b}{\delta r_D} & = &  0.6744~{\rm fm}\,.
\label{eq18}
\eeqa

In the second case, only the experimental value used for $Q$ is changed to 
obtain $\delta b/\delta Q$. The latest two published measurements of the 
experimental 
value of the deuteron quadrupole moment $Q$ are the  value of Reid and Vaida 
\cite{44}, $Q=0.2860\pm$$0.0015$ fm$^2$, and the value of Bishop and Cheung \cite{32}, 
$Q=0.2859\pm0.0003$ fm$^2$. To study the correlation between the extracted 
value of $b$ and $Q$ we used the 
experimental value $Q=0.2860\pm0.0015$ fm$^2$ \cite{44}. The result of the analysis 
is 
\beqa
b & = & 0.0205\pm0.0011~{\rm fm}^2\,.
\label{eq19}
\eeqa
From these calculations we show that the increase in the value of $Q$ leads 
to a small decrease in the value of $b$ resulting in 
\beqa
\frac{\delta b}{\delta Q} & = & -0.0052\,.
\label{eq20}
\eeqa
Similarly, we found
\beqa
\frac{\delta b}{\delta A_S} & = & -0.098~{\rm fm}^{3/2}\,,
\label{eq21}
\eeqa
and
\beqa
\frac{\delta b}{\delta \eta} & = & -0.002~{\rm fm}^2\,.
\label{eq22}
\eeqa
From these results we conclude that the correlation between $b$ and $r_D$ is 
the dominant one.

\begin{figure}[htb]
\begin{center}
\includegraphics[scale=0.7]{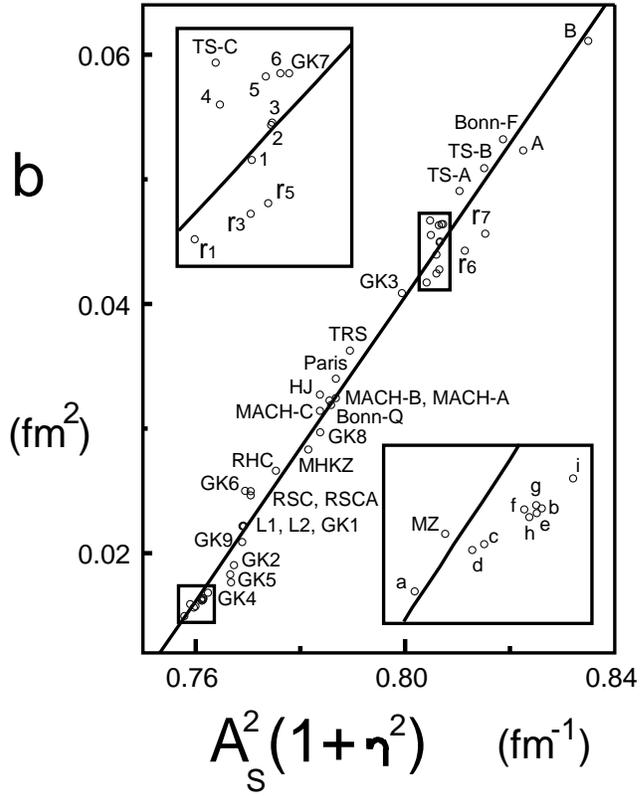}
\vspace*{-4.5cm}
\caption{The variation of the slope $b$ versus $A_S^2(1+\eta ^2)$ of standard local
potential models. The middle (lower) part of the graph 
is magnified in the upper (lower) inner frame. The value of $b$ of Eq.~(\ref{eq13}) 
is extracted from this straight line corresponding to 
$A_S^2(1+\eta ^2)=0.7817$ fm$^{-1}$.}
\label{fig1}
\end{center}
\end{figure}

\begin{figure}[htb]
\begin{center}
\includegraphics[scale=.7]{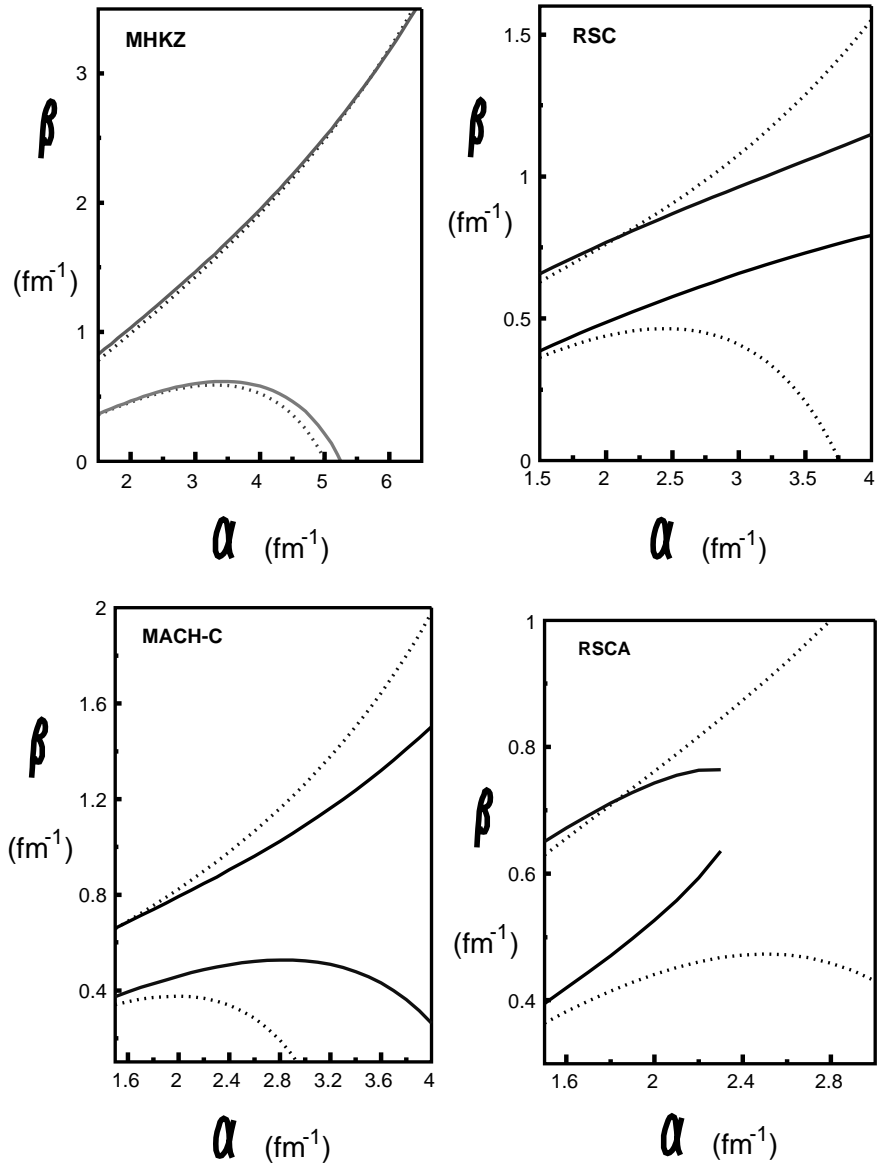}
\vspace*{-0.5cm}
\caption{The variation of the "strength" parameter $\beta$ versus the "range" parameter $\alpha$ for some local potential models, 
indicated on the graphs, which have intersection points, i.e., have a 
pair ($\alpha,\beta$) which give the experimental values of both $Q$ and $r_D$. The solid (dotted) lines represent the values of $\alpha$ and $\beta$ 
which give the experimental values of $Q$ ($r_D$).}
\label{fig2}
\end{center}
\end{figure}

\begin{figure}[htb]
\begin{center}
\includegraphics[scale=.7]{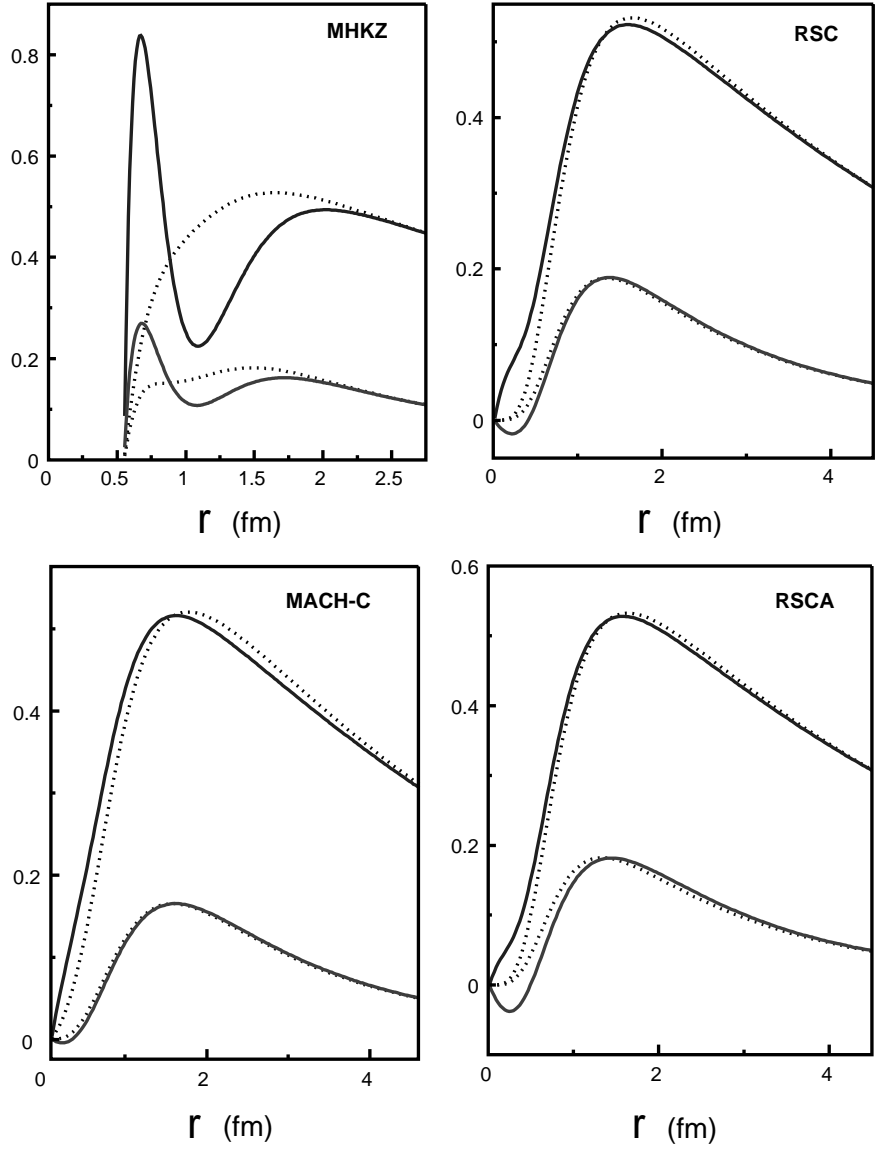}
\vspace*{-0.8cm}
\caption{Transformed wave functions having the experimental values of 
both $r_D$ and $Q$ produced by the unitary transformation 
(solid line) are compared to the local potentials which are represented by the dotted line. The upper (lower) curves are the $u$ $(w)$ wave 
functions.}
\label{fig3}
\end{center}
\end{figure}

\begin{figure}[htb]
\begin{center}
\includegraphics[scale=.7]{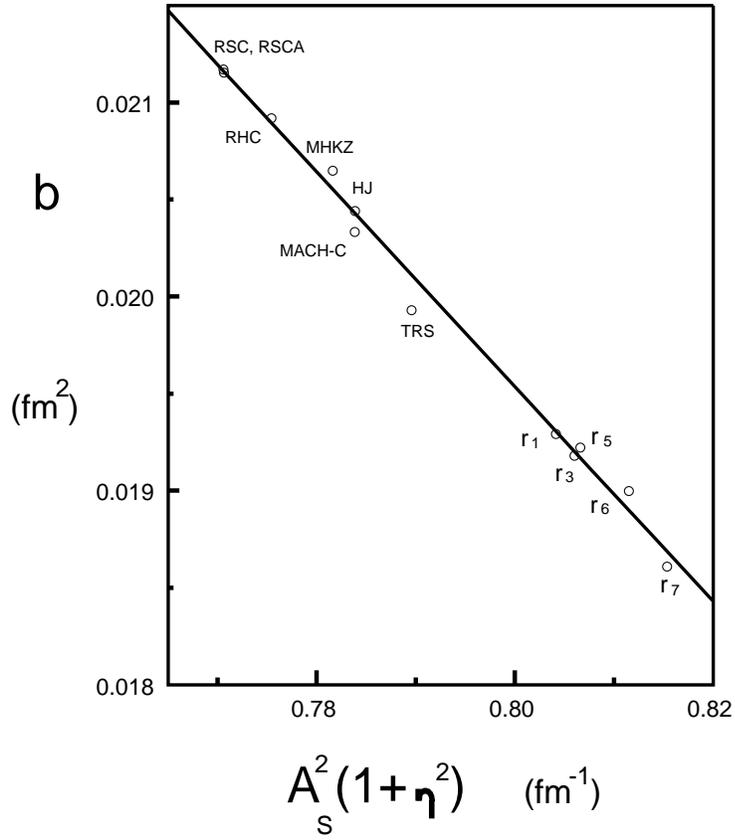}
\vspace*{-4cm}
\caption{The correlation of the slope $b$ versus $A_S^2(1+\eta ^2)$ of the twelve nonlocal
"transformed" potential models having the experimental values of 
$r_D$ and $Q$. The new value of $b$ of Eq.~(\ref{eq15}) is 
extracted from this line. It is the value of $b$ corresponding to the experimental value of $A_S^2(1+\eta^2)=0.7817$ fm$^{-1}$.}
\label{fig4}
\end{center}
\end{figure}

Now, we would like to point out the reason for the changing slope of the 
correlation between $b$ and $A_S^2(1+\eta^2)$ from being positive in the 
case of the local potential models of Fig.~\ref{fig1} to a negative value in the case 
of the nonlocal potential models of Fig.~\ref{fig4}. Since $b$ is proportional 
to $r_D$ and $r_D$ in turn proportional to $A_S^2(1+\eta^2)$, 
the value of $b$ is proportional to $A_S^2(1+\eta^2)$. The values of $A_S$ 
and $\eta$ and then $A_S^2(1+\eta^2)$ are the same for the local and the 
corresponding nonlocal model, but $b$ and $r_D$ are different. 
We see that the value of $b$ in the case of using the radial 
wave functions of the local model is considerably greater than that in the 
case of using the transformed wave functions of the corresponding nonlocal 
potential model. Therefore, the slope will change sign in the case of using 
the transformed wave functions of the nonlocal potential models.

\begin{ack}
E.M.\ Darwish would like to thank H.\ Arenh\"ovel as well as his theory group 
for the fruitful discussions and reading the manuscript.
\end{ack}


\newpage

\begin{table}[ht]
\begin{center}
\caption{The experimental values used in the analysis of this work.}
\vspace*{0.3cm}
\begin{tabular}{ll}\hline\hline
      Value                        & Ref. \\ \hline \vspace*{-0.2cm}
$r_D$=1.9547$\pm$0.0019 {\rm fm}          & \cite{9}   \\ \vspace*{-0.2cm}
$Q$= 0.2859$\pm$0.0003 {\rm fm}$^2$      & \cite{32}  \\ \vspace*{-0.2cm}
$A_S$=0.8838$\pm$0.0004 {\rm fm}$^{-1/2}$ & \cite{33}  \\ \vspace*{-0.2cm}
$\eta$=0.02713$\pm$0.00006                   & \cite{30}   \\ 
$E_b$= -2.224575$\pm$0.000009 {\rm MeV}    & \cite{34}   \\ 
\hline \hline
\end{tabular}
\label{tab1}
\end{center}
\end{table}

\vspace*{1cm}

\begin{table}[ht]
\begin{center}
\caption{The pairs of the nonlocality parameters ($\alpha,\beta$) 
of the unitary transformations producing transformed wave functions 
having the experimental values of both $r_D$ and $Q$.}
\vspace*{0.3cm}
\begin{tabular}{lccc}\hline\hline
local   & Ref.&$\alpha$ & $\beta$     \\
potentials & &fm$^{-1}$& fm$^{-1}$ \\ \hline
MHKZ    & \cite{4} &5.70890&   2.9672  \\
GK2     & \cite{19} &1.5818&   0.3966  \\
GK3     & \cite{19} &2.0516&   0.9938  \\
GK5     & \cite{19} &1.6421&   0.4143  \\
GK7     & \cite{19} &2.0455&   1.0127  \\
GK8     & \cite{19} &2.0894&   1.0066  \\
RSC     & \cite{21} &2.0650&   0.7787   \\
RHC     & \cite{21} &2.7131&   1.2650   \\
RSCA    & \cite{21} &1.8211&   0.7141   \\
MACH-C  & \cite{22} &1.5363&   0.6680   \\
r1      & \cite{29} &2.2160&   0.9669   \\
r3      & \cite{29} &2.3759&   1.1125   \\
r5      & \cite{29} &2.9344&   1.4908   \\
r6      & \cite{29} &3.5364&   1.8926   \\
r7      & \cite{29} &3.0106&   1.6228   \\
TRS     & \cite{25} &1.4821&   0.6472   \\
HJ      & \cite{26} &3.0968&   1.4556   \\
\hline \hline
\end{tabular}
\label{tab2}
\end{center}
\end{table}

\end{document}